\documentclass[preprint]{elsarticle}

\usepackage{amssymb}
\usepackage{subfigure}
\usepackage{lineno}
\journal{Nuclear Instruments and Methods}

\begin{document}
\begin{frontmatter}
\title{An LED-based Flasher System for VERITAS}
\author{D. Hanna, A. McCann, M. McCutcheon and L. Nikkinen}
\address{Department of Physics,
McGill University,
Montreal, QC H3A2T8, Canada}

\begin{abstract}
We describe a flasher system designed for use in monitoring the gains of the 
photomultiplier tubes used in the VERITAS gamma-ray telescopes.
This system uses blue light-emitting diodes (LEDs) so it 
can be operated at much higher rates than a traditional laser-based system. 
Calibration information can be obtained with better statistical precision
with reduced loss of observing time.
The LEDs are also much less expensive than a laser. 
The design features of the new system are presented, along with measurements 
made with a prototype mounted on one of the VERITAS telescopes.

\end{abstract}
\end{frontmatter}

\section{Introduction}

This article concerns a light flasher system developed for the Very Energetic 
Radiation Imaging Telescope Array System (VERITAS)~\cite{weekes02, holder06}.
VERITAS is an array of four imaging atmospheric Cherenkov telescopes (IACTs)
located at the Fred Lawrence Whipple Observatory 
on Mount Hopkins in southern Arizona. 
The VERITAS telescopes measure the Cherenkov light caused by relativistic 
electrons in air showers initiated by high energy astrophysical gamma rays.
Each telescope consists of a 12-m-diameter reflector 
viewed by a `camera' comprising 
499 29-mm-diameter photomultiplier tubes (PMTs).
The PMTs are read out using FADCs operating at rate of 500 megasamples per
second.
A flasher system is used to monitor the gains and timing characteristics of 
the PMTs.

The present VERITAS flasher system~\cite{hanna07} 
uses a nitrogen laser which produces 
short pulses of ultraviolet ($\lambda$ = 330 nm) light that are distributed
through quartz optical fibres.
Each telescope is supplied by one fibre, which ends approximately four metres 
from the front of the camera in a termination box equipped with an opal
diffuser.
This system provides uniform and simultaneous illumination to all PMTs 
in the camera.
It delivers light pulses which are similar  
in shape and wavelength to the Cherenkov-light pulses seen in normal 
observing and is therefore well-suited for testing the PMTs in their
nominal region of operation.

This system has a few disadvantages.
It cannot be pulsed at a rate much higher than 10 Hz which 
means that calibration runs can be long, especially when one is 
obtaining high-statistics data sets.
Such runs are usually performed under dark-sky conditions and are 
therefore at the expense of scientific observing time.
Another negative feature of the laser-based system is the cost and 
lifetime of the laser which uses a factory-sealed cartridge capable of
producing a large but finite number of pulses. 
Replacing the cartridge costs several thousand dollars and can cause
considerable down time if a spare is not on hand.

To overcome these disadvantages, we have developed a flasher system based on 
blue light-emitting diodes (LEDs) which are bright enough and fast enough to 
illuminate the VERITAS PMTs with pulses of duration and 
intensity that are very similar to those coming from the laser-based system
and yet cost less than a dollar each. 
The LED flasher can be used to test the linearity of the PMT-FADC 
chain and can be used in conjunction with a neutral-density 
filter to determine the single-photoelectron response in all the channels.
Moreover, because the system can be flashed at rates limited only by the
VERITAS data acquisition system, runs which require very high statistics 
can be carried out in only a few minutes.

\section{Flasher Design}

\begin{figure}
\centering
\subfigure
{
\includegraphics[width=0.45\textwidth]{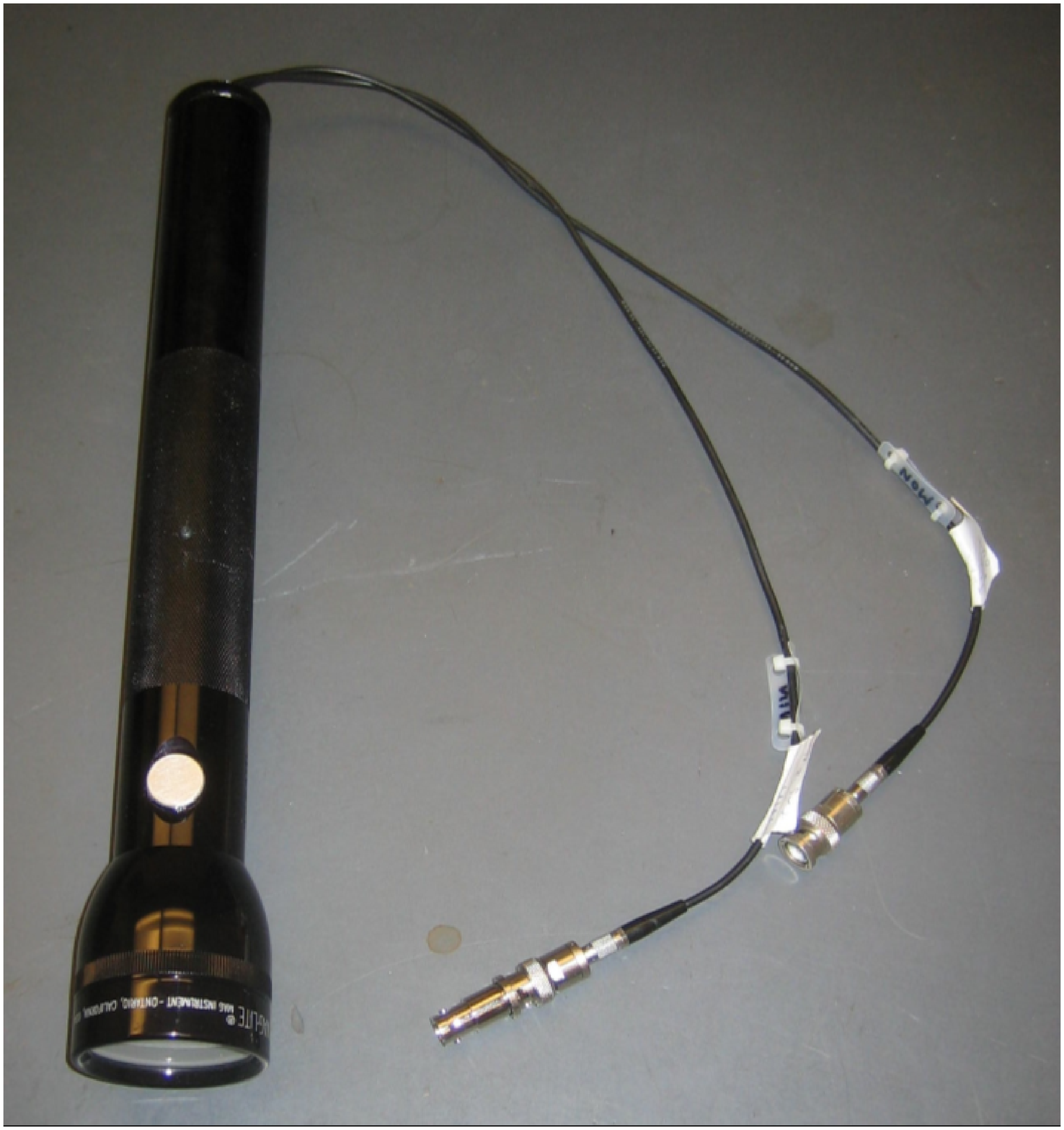}
}
\subfigure
{
\includegraphics[width=0.45\textwidth]{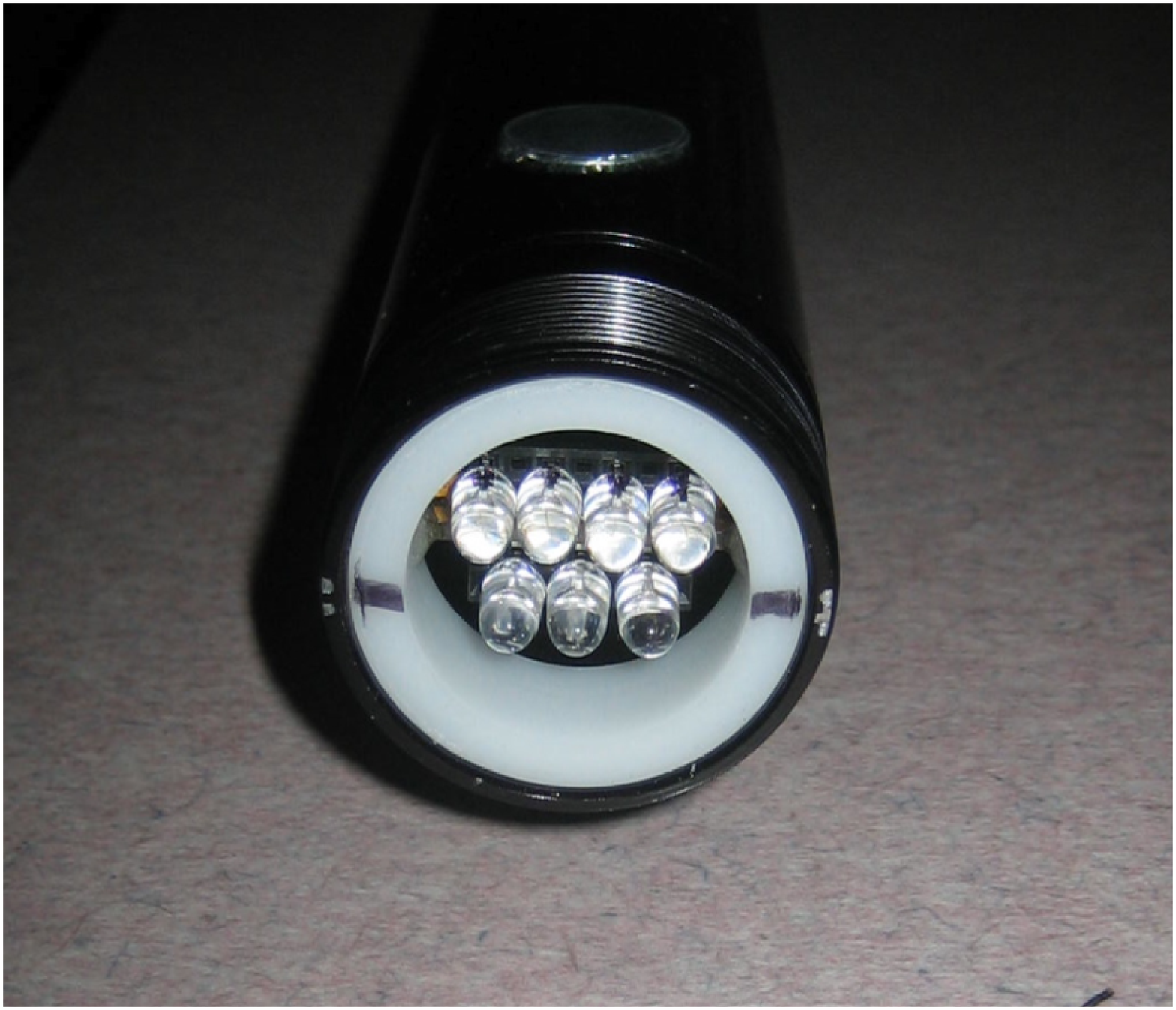}
}
\caption{ Left: A photograph of the assembled LED-based flasher. 
The system is housed 
in a rugged Maglite$^{\tiny\textcircled {R}}$  
container and comprises seven blue LEDs, driver 
electronics and batteries. The front face of the device is an opal diffuser and 
the flying leads are for provision of an external trigger in and a 
trigger-synch pulse out.
Right: The LED cluster as seen with the diffuser assembly removed.
The seven LEDs are chosen to have approximately equal output so that when they 
are pulsed in a sequence of suitable combinations the result is an intensity 
`ramp'.
}
\label{flasher}
\end{figure}

The LED flasher assembly is pictured in Figure~\ref{flasher}.
The LEDs and associated electronics are housed in a modified 
Maglite$^{\tiny\textcircled {R}}$ which 
provides protection from weather and dust allowing the device to be mounted 
on the telescope for extended periods.
At the front, a 50-mm opal diffuser (Edmund Optics NT46-106) replaces the 
lens of the Maglite$^{\tiny\textcircled {R}}$  and spreads the light
from the LEDs, located a few mm behind it, such that each PMT in the camera
receives approximately the same amount of light.
Two cables emerge from the rear of the flasher. One is used for supplying an
external NIM-level pulse to activate the flasher and the other is used for a
TTL-level trigger-out signal. 

\begin{figure}[]
\centerline{\includegraphics[angle=90.,width=1.0\textwidth]{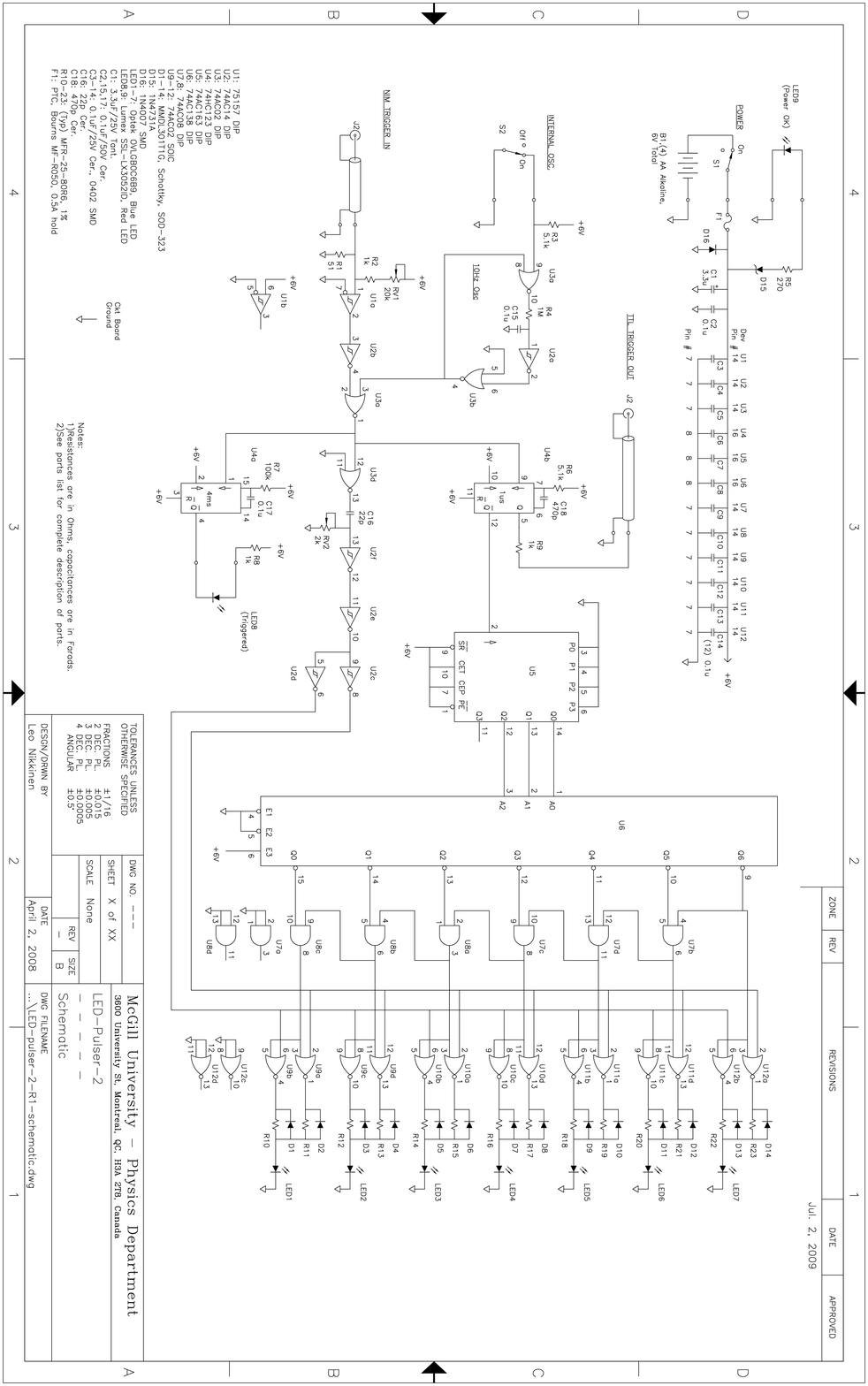}}
\vspace*{0.0cm}
\caption{The schematic diagram for the flasher electronics. The circuit is 
driven by pulses from an external source or an internal oscillator.
These pulses increment a binary counter (U5)  which cycles repeatedly through 
the states from 000 to 111. The output bits are sent to a decoder (U6) with 
active-low outputs coupled to a set of AND gates (U7 and U8). 
The output of these are used to drive the LEDs using NOR gates (U9 and U12) 
to source the current required.
}
\label{schematic}
\end{figure}

The blue LEDs can be seen in Figure~\ref{flasher} which is a photograph taken 
with the diffuser removed.
These LEDs (Optek OVLGB0C6B9) are InGaN devices with a peak wavelength of 
approximately 465 nm, nominal brightness of 3200 mcd, and a 50\% power
angle of 6$^\circ$.

The flasher circuit is shown in Figure~\ref{schematic} and is based on concepts
described in~\cite{patterson}.
It uses simple integrated circuits to source currents through the blue LEDs 
each time a pulse is received, either from an external 
source or from an internal 10 Hz oscillator.
These pulses increment a binary counter (U5)  which cycles repeatedly through 
the states from 000 to 111. The output bits are sent to a decoder (U6) with 
active-low outputs coupled to a set of AND gates (U7 and U8). 
The output of these are used to drive the LEDs using NOR gates (U9 and U12) 
to source the current required.
The logic is arranged such that the number of illuminated LEDs in the array
runs from zero through seven sequentially so that over the course of many 
cycles the light intensity is ramped from off to maximum repeatedly. 
This is useful for scanning over the range of the PMT and FADC response.

The width of the current-pulse is adjustable and is made narrow enough to 
achieve a light pulse from the LEDs which is a few nanoseconds in length.
This limits the amount of light in the pulse 
since the slew rates in the electronics 
are such that full amplitude is not attained before the pulse is turned off.

\section{On-site Tests with a VERITAS Camera}  

The LED flasher was tested using one of the VERITAS telescopes and the standard
data acquisition system. 
For purposes of comparison, data were acquired using the laser system 
immediately afterwards.
The LED flasher rate was 200 Hz and the laser was pulsed at 10 Hz.
For each event a 24-sample (48 ns) FADC trace for each PMT was written to a data
file. 
Pedestal events were included in the data stream at the rate of 1 Hz.
 
\subsection{Pulse Shape}

We have attempted to produce with the LED system 
a pulse shape which is similar to that produced by
the laser system, which is itself meant to be narrow, like the pulses
caused by the Cherenkov photons from air showers.

\begin{figure}[]
\centerline{\includegraphics[width=1.0\textwidth]{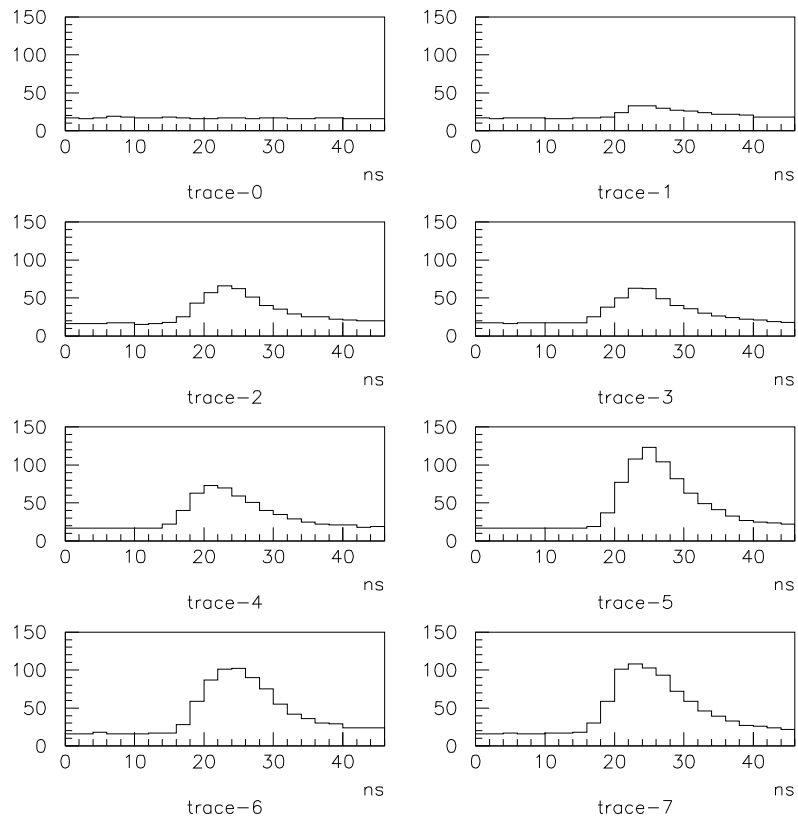}}
\vspace*{0.0cm}
\caption{
FADC traces for eight sequential flashes of the LED system as seen by one PMT. 
The sample size in arbitrary units is plotted against 
sample time in nanoseconds.
With each panel, the number of LEDs contributing to the flash is
incremented by one.
}
\label{traces}
\end{figure}

Figure~\ref{traces} shows an example of the pulses obtained from one PMT 
over the course of an eight-event cycle. 
The light level increases from zero, with no LEDs on, to maximum with seven LEDs
on. 
Note that due to pulse-to-pulse fluctuations and differences in the mean 
output of each LED, the traces do not follow a strictly monotonic
increase with event number in this example.

\begin{figure}[]
\centerline{\includegraphics[width=1.0\textwidth]{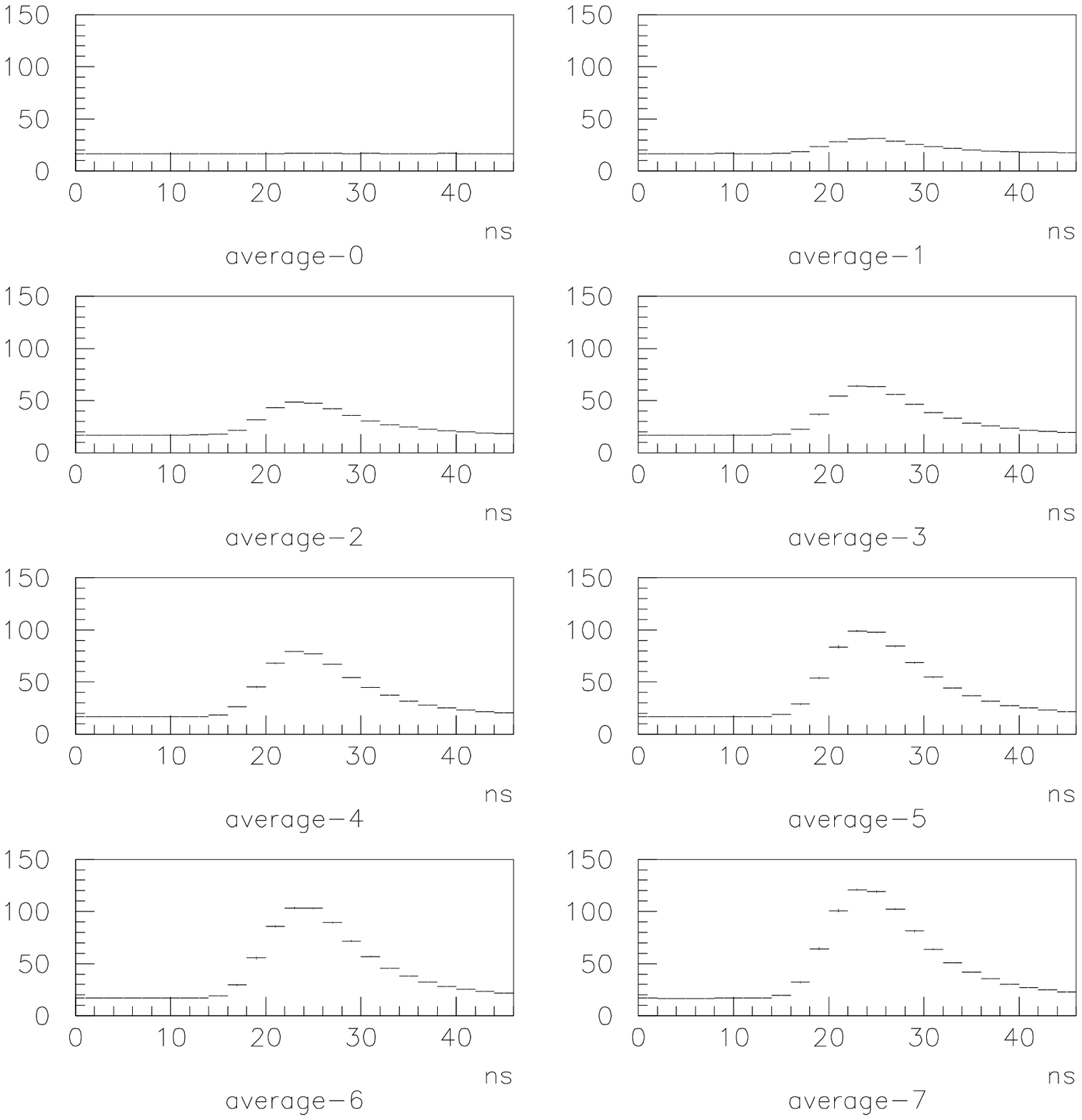}}
\vspace*{0.0cm}
\caption{
As in Figure~\ref{traces} but averaged over 250 LED pulses at each 
intensity level.
}
\label{average}
\end{figure}

The averages of 250 events are shown in Figure~\ref{average} where the progress
is more linear except for level 7 which is very similar to level 6, indicating 
a weak sixth LED.

\begin{figure}[]
\centerline{\includegraphics[width=1.0\textwidth]{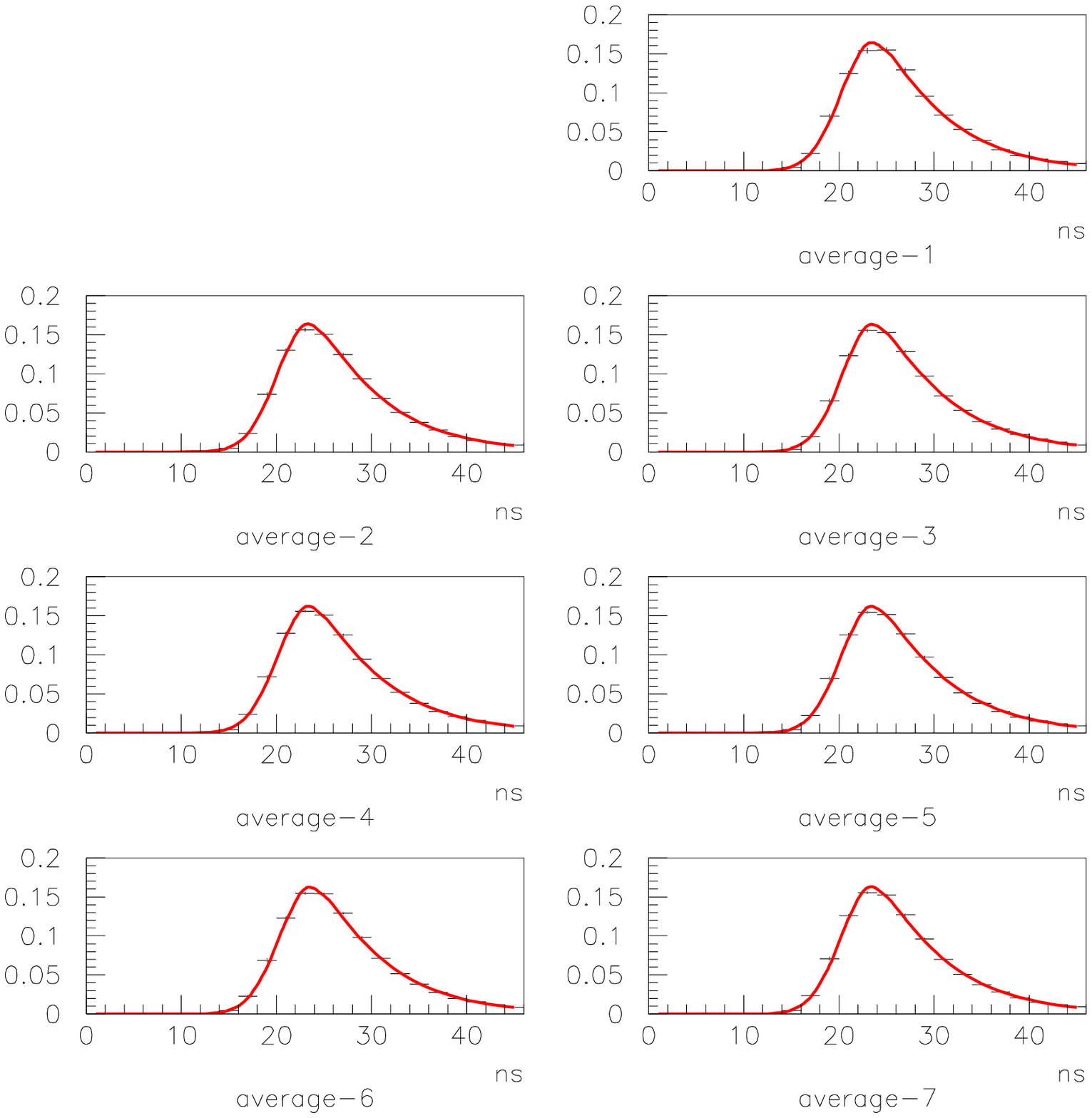}}
\vspace*{0.0cm}
\caption{
As in Figure~\ref{average} but the distributions have been normalized 
to 1.0 and fit with an asymmetric Gaussian described in the text.
}
\label{fits}
\end{figure}

In Figure~\ref{fits} the distributions of Figure~\ref{average} have been 
normalized to 1.0 and fit with an asymmetric Gaussian function of the form

\[ Q = Q_0~e^{-(t-t_0)^2/(2\sigma^2)}~~~   t < t_0 \] 

\[ Q = Q_0~e^{-(t-t_0)^2/(2\sigma^2+\alpha(t-t_0))}~~~   t > t_0 \]

The fit parameters are independent, within statistical uncertainties, of the 
the number of LEDs contributing to the flash.

\begin{figure}[]
\centering
\subfigure
{
\includegraphics[width=0.45\textwidth]{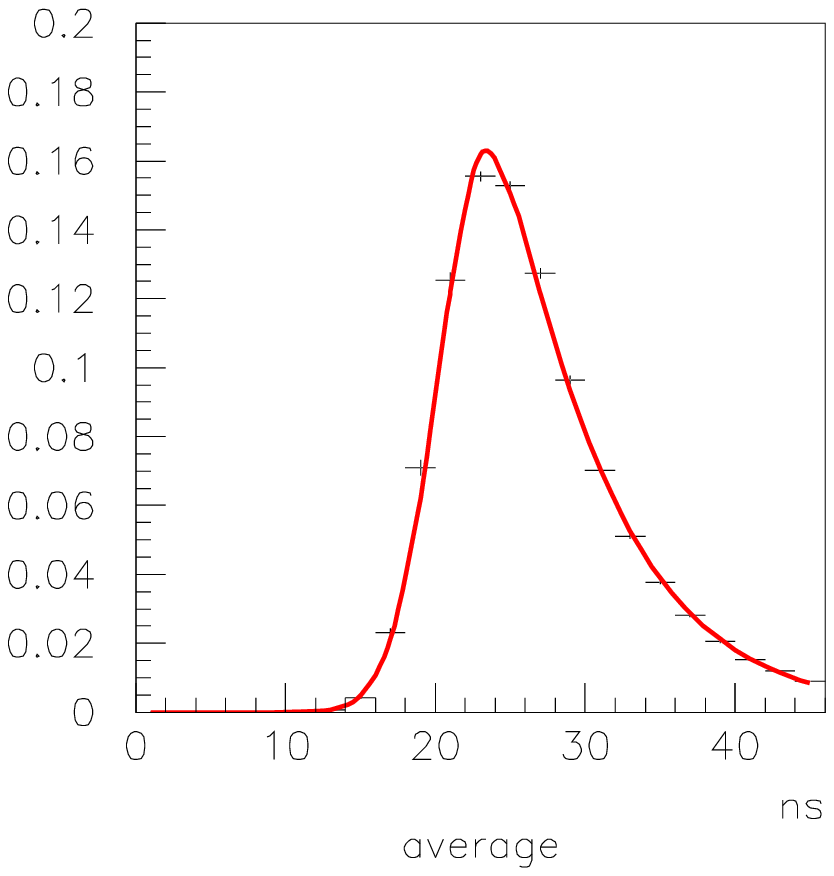}
}
\subfigure
{
\includegraphics[width=0.45\textwidth]{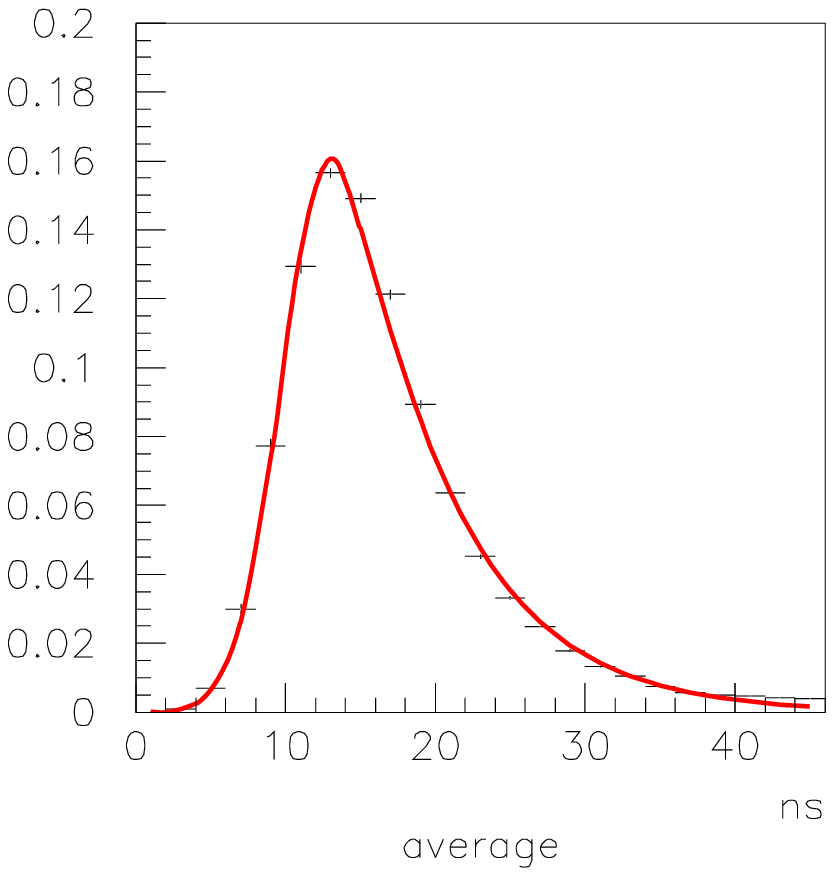}
}
\caption{
Left: Normalized average trace and fit for the LED data from events where 
all seven LEDs were
pulsed simultaneously.
Right:Normalized average trace and fit for laser data.
}
\label{example}
\end{figure}

Figure~\ref{example} (left) is a repeat, in larger format, of the `average-7'
panel in Figure~\ref{fits} and Figure~\ref{example} (right) is a similar plot 
made using laser data\footnote{The laser has a single intensity so a series 
of eight plots like in 
Figure~\ref{fits} is not possible.}.
The two figures look similar and, except for $t_0$, the fit parameters are 
identical within statistical uncertainties.
Thus the LED flasher performs as well as the laser flasher at
simulating Cherenkov pulses.

\subsection{Pulse Stability} 

In this section we compare the fluctuations of the LED flasher with those
of the laser. 
The LED-based system does not include an independent monitor so we use 
the `camera average' for measuring light levels. 
The charges reported by all 499 channels in the camera are combined for each
laser or LED pulse. 
A few dead or miscreant channels are thereby included but do not 
influence the average significantly. 
Combining charges from the entire camera also means that fluctuations from 
finite photostatistics in individual channels will be made negligible.

\begin{figure}[]
\centerline{\includegraphics[width=1.0\textwidth]{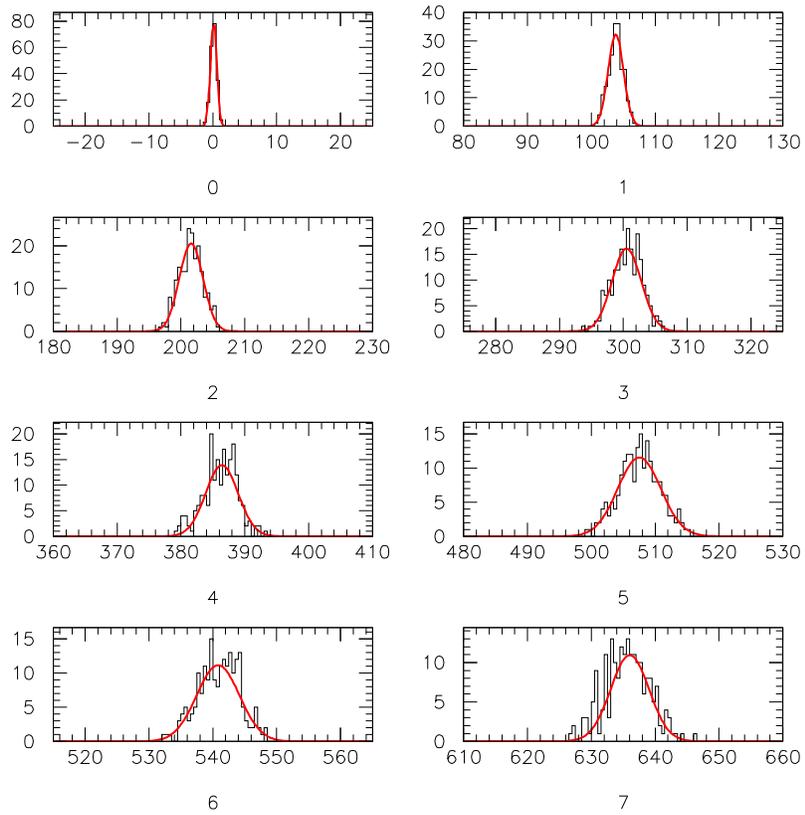}}
\vspace*{0.0cm}
\caption{
Distributions of the average pulse size (in arbitrary units)
for eight different LED levels.
The averages are made using all PMTs in the camera.
}
\label{led_res_1}
\end{figure}

\begin{figure}[]
\centerline{\includegraphics[width=1.0\textwidth]{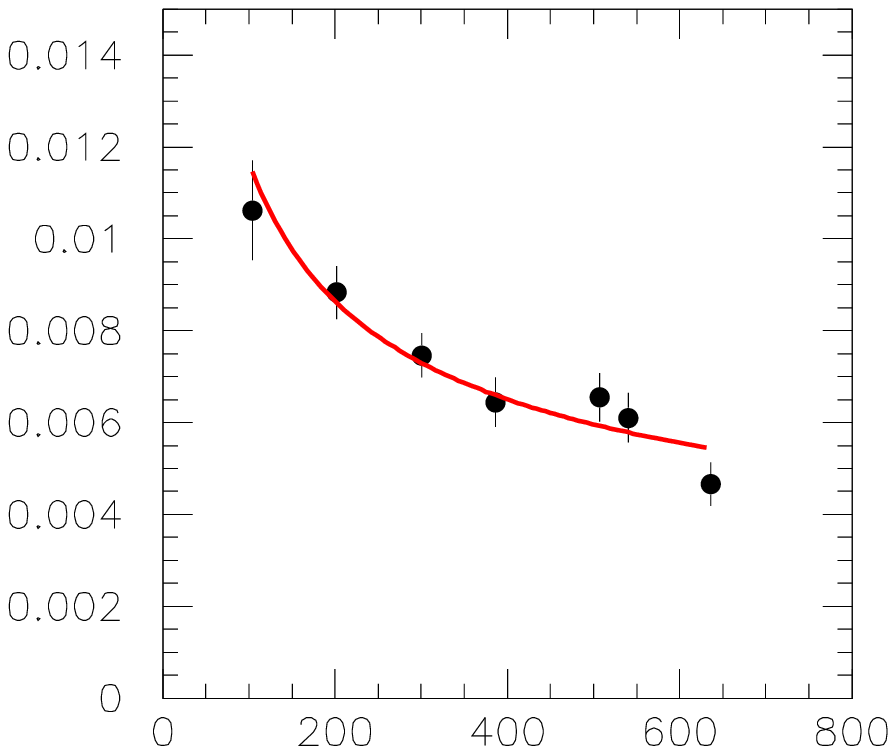}}
\vspace*{0.0cm}
\caption{
Plot of resolution ($\sigma/\mu$)  vs mean ($\mu$) (in arbitrary units)
of the 
distributions in Figure~\ref{led_res_1}. 
The curve is of the form $\sigma/\mu = A + B/\sqrt{\mu}$, expected when 
combining multiple, identical but uncorrelated, sources.
}
\label{led_resolution}
\end{figure}

Figure~\ref{led_res_1} shows the distributions of camera averages for the eight
(including zero) LED light levels together with Gaussian fits. 
The fitted values are displayed in Figure~\ref{led_resolution} 
(left) where the resolution 
($\sigma/\mu$) for each non-zero light level is plotted against the 
corresponding mean ($\mu$).
For multiple, identical but uncorrelated, sources one expects the data to 
behave like $\sigma/\mu = A + B/\sqrt \mu$, which is the form of the 
function that has been fit to the data in Figure~\ref{led_resolution}.

\begin{figure}[]
\centerline{\includegraphics[width=1.0\textwidth]{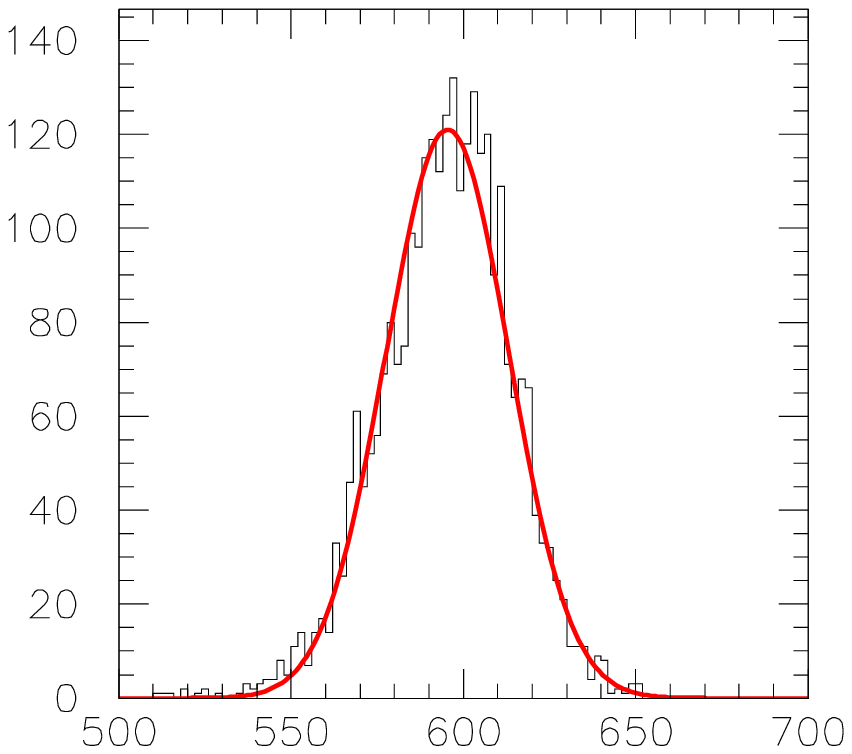}}
\vspace*{0.0cm}
\caption{
Plot of the camera-average response (in arbitrary units)
to pulses from the laser at its nominal intensity.
The resolution ($\sigma/\mu$) is approximately 3\%.
}
\label{laser_resolution}
\end{figure}

The laser run used for these studies has only one intensity level. 
The distribution of camera averages is displayed in 
Figure~\ref{laser_resolution}  
along with a Gaussian fit that shows the resolution to be approximately 3\%.
Note that at a similar light level ($\mu \simeq 600$) the LED system has a 
resolution of 0.6\%.
This results from the good intrinsic resolution of each LED, coupled with 
the statistical improvements due to combining multiple LEDs.

\section{PMT Calibration}

In this section we will show how the LED
flasher can be used to monitor the PMT 
gains and we will demonstrate that it performs at least 
as well as the laser-based
system in this regard.

\subsection{Relative Gains and Flat-fielding}

The simplest way to measure the relative 
gains of PMTs is to provide light 
flashes to the camera and plot the response of each PMT against the response of 
some kind of monitor which measures the absolute light output of the flasher.
In the case of the laser, a photodiode measures a fraction of the laser light 
and, since the laser is nominally a constant intensity device, the relative 
gain of the PMT is computed as the ratio of the mean pedestal-subtracted 
PMT pulse size to the mean monitor pulse size. 
This assumes, but does not demonstrate, linearity in the system.

\begin{figure}[]
\centerline{\includegraphics[width=1.0\textwidth]{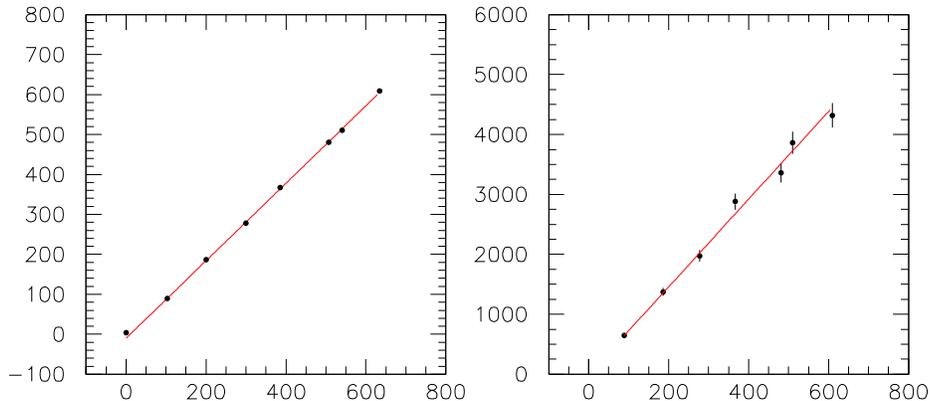}}
\vspace*{0.0cm}
\caption{Plots to illustrate the use of the LED flasher in 
calibration.
The left-hand plot shows the mean pulse size (in arbitrary units) from a 
typical PMT vs the mean value (in arbitrary units) of the monitor 
(average over all channels in the camera) for eight light levels. 
The fitted line has a slope which is proportional to the gain of the PMT,
its quantum efficiency and the relative amount of light it receives from the 
flasher.
The right-hand plot shows the variance ($\sigma^2$) of the charge 
distribution for each light level vs the mean value of the charge distribution 
for that PMT. 
Coherent fluctuations of the flasher have been removed using the monitor, and 
pedestal fluctuations have also been unfolded.   
}
\label{led_slopes}
\end{figure}

Using the camera-average monitor introduced previously, 
plots like those seen in Figure~\ref{led_slopes} can be constructed.
In the left-hand plot the mean charge of a typical PMT is plotted against 
the mean 
monitor for the eight (including zero) light levels produced by the LED system.
A straight line has been fit to the data.
For a given channel the slope of this line is related to the relative amount of
light received by the PMT, the quantum efficiency of the PMT, its efficiency
at collecting photoelectrons from the photocathode onto the first dynode, and 
the gain of the subsequent dynodes, pre-amplifier, etc.
A change in any of these components will cause the slope to change and if 
one of the components changes (eg the quantum 
efficiency decreases) the dynode gain can be changed to compensate.
Measuring the slopes periodically and providing a list of high-voltage 
adjustments to equalize them (flat-fielding) is the primary 
task of the gain-monitoring system.

Note that the eight points demonstrate the linearity of the system.
If it can be assumed that linearity prevails, multiple points are not 
necessary; 
with a pedestal and only one non-zero point one can compute the slope.
This is the procedure followed when using the laser system; 
with this LED system linearity can be proven rather than assumed.
It should be stated that in the foregoing we are assuming implicitly that 
the monitor made from an average of many channels has linear behaviour.

\begin{figure}[]
\centerline{\includegraphics[width=1.0\textwidth]{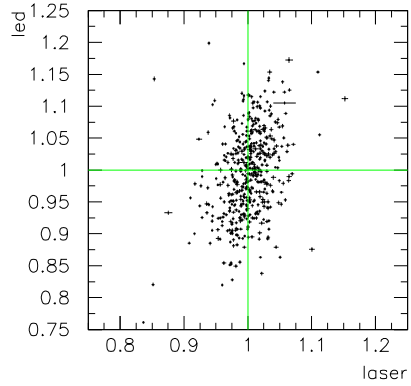}}
\vspace*{0.0cm}
\caption{Comparison of relative gains determined with the laser and the 
LED flashers for all channels in the camera.
The slope of mean pulse size vs monitor for each channel, determined 
with LED data, is
plotted vs the mean pulse size for each channel determined with laser data.
Both data sets have been scaled by their mean value.
The laser data are more strongly peaked ($\sigma/\mu = 2.8\%$) than the LED
data ($\sigma/\mu = 6.4\%$) since they are used in the flat-fielding procedure. 
}
\label{led_laser_3}
\end{figure}

It is instructive to compare the relative gains determined using 
the LED flasher with those determined using the laser. 
Those determined with the laser are all clustered about 
a single mean value to within errors. This is by construction since relative 
gain is the parameter used for flat fielding.
The relative gains determined using  the LED flasher would also cluster 
about that value
if the LED light had the same wavelength as that of the laser.
However, since the wavelength is longer (470 nm vs 330 nm) we expect
differences due to the wavelength-dependent quantum efficiency of the 
photo-cathodes. To the extent that this dependence varies from PMT to PMT we
expect any clustering to be broadened.
This phenomenon can be seen in Figure~\ref{led_laser_3}.

\subsection{Absolute Gains}

\subsubsection{Photostatistics}

The right-hand plot in Figure~\ref{led_slopes}
shows the variance of the PMT's pulse size distribution 
vs its mean for the different light levels. The contribution of pedestal 
fluctuations and pulse-to-pulse fluctuations of the LEDs have been unfolded.
If we say that only fluctuations due to the Poisson statistics of 
photoelectron production remain, we have the following argument:
the mean number of photoelectrons hitting the first dynode for a given light 
level is $N_{pe}$.
There are fluctuations about this given by (approximately Gaussian) 
Poisson statistics $\sigma_{pe} \simeq \sqrt{N_{pe}}$.
After the dynodes and preamplifier have amplified this 
we have $\mu = G N_{pe}$ and $\sigma = G \sqrt{N_{pe}}$.
and we expect $\sigma^2 = G^2 N_{pe} = G \mu $.
Thus the slope of the right-hand plot is related to the net gain 
after the first dynode.
Note that this method 
provides an estimate of the absolute gain without the use of a 
calibrated monitor to indicate the light level.
The camera-average monitor is only used to remove 
pulse-to-pulse variations in light output from the LEDs (a small effect
anyway, as demonstrated earlier).

A final but important effect has been left out in this simple treatment.
Taking into account statistics at the other dynodes, which are in general 
described by a Polya distribution~\cite{prescott}, leads to a correction 
factor such that $G = \sigma^2/\mu/(1+\alpha^2)$ where $\alpha$ 
is the resolution  
of the distribution that would result from injecting only 
single photoelectrons into the dynode chain. 
The effect of this is shown later in this article.

\begin{figure}[]
\centerline{\includegraphics[width=1.0\textwidth]{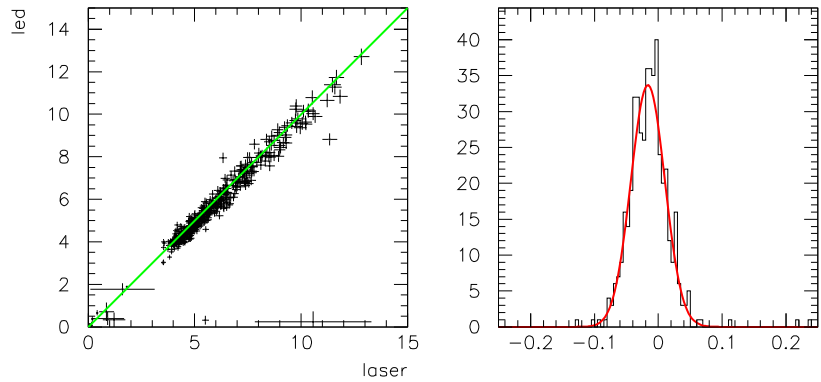}}
\vspace*{0.0cm}
\caption{Comparison of gains from photostatistics made with the laser and the 
LED flashers for all channels in the camera.
In the left plot the gains determined using the LED flasher are plotted against 
those determined using the laser.
The scales are in arbitrary units. 
The solid line is to guide the eye and is not a fit.
In the right plot, the difference divided by the sum of the two measurements
is histogrammed. 
The distribution has a mean of -1.6 \% and a sigma of 2.7 \%. 
}
\label{led_laser}
\end{figure}

Note that the absolute gains determined in this way do not include effects of
the photocathode quantum efficiency and the first-dynode collection efficiency.
Thus we expect a better correlation between measurements made with the LED 
system and measurements made with the laser system.
This is seen in Figure~\ref{led_laser} where gains determined using the LED 
system are plotted against those determined using the laser system.

\subsubsection{Single Photoelectrons}

A very direct way to measure the gain of a PMT is to determine the mean 
value of the pulse size distribution due to single photoelectrons.
A single photoelectron should, on average, result in a measured charge of 
$Ge$ Coulombs where $G$ is the system gain and $e$ is the electron charge, 
$1.6 \times 10^{-19}$ Coulombs.
In this case, as with the photostatistics 
method, the `gain' includes everything 
from the first dynode through to the digitized data.
The photocathode quantum efficiency and first-dynode collection 
efficiency are not included.

Single photoelectrons are measured in VERITAS by 
taking special laser runs with a 
neutral-density filter\footnote{This `filter' is a thin sheet of aluminum with 
a small hole aligned with the centre of each PMT.
See~\cite{hanna07} for details.} 
covering the PMTs to attenuate spurious light from the night sky.
The light level from the laser is adjusted to  
result in the release of a single photoelectron in some events.
With a low enough laser intensity, one obtains a distribution with mostly 
zero photoelectrons, a reasonable fraction of single photoelectrons and 
smaller fractions of two, three, etc photoelectrons, following 
a Poisson distribution.
Assuming that each photoelectron gives rise to a Gaussian distribution after
the effects of amplification and that this distribution is convoluted with 
another Gaussian distribution, which results from the effect of electronic noise
(ie the pedestal distribution), the entire distribution can be fit with 
only five parameters.
These are: the pedestal mean and width, the single-photoelectron mean and
width, and the mean number of photoelectrons.

This technique has become a standard procedure using the 
VERITAS laser system but it has limitations.
The technique requires a small mean number of
photoelectrons, so most laser pulses contribute to the 
zero photoelectron peak (ie the pedestal).
This means that many pulses are needed and 
since the laser can only be pulsed at a rate of approximately 10 Hz, 
very long (order one hour) runs under dark skies are required to attain
required levels of precision;
this results in a significant loss of good observing time.
Another issue is that of laser lifetime.
Lasers of the type used in the VERITAS system can only deliver a finite 
number of pulses before needing a new, expensive, cartridge.

The LED flasher was designed in part to address these problems.
Its flash rate is limited only by the data acquisition system 
(which can handle rates of up to 400 Hz) 
and it is  based on readily available and remarkably low-cost (less than one 
dollar each) LEDs.

\begin{figure}[]
\centering
\subfigure
{
\includegraphics[width=0.47\textwidth]{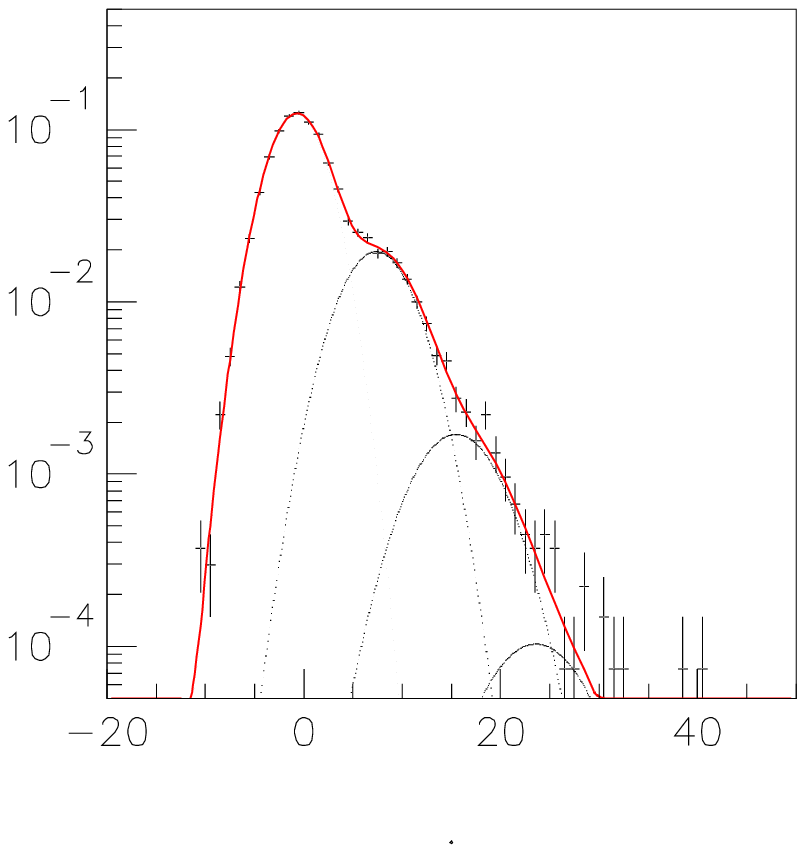}
}
\subfigure
{
\includegraphics[width=0.47\textwidth]{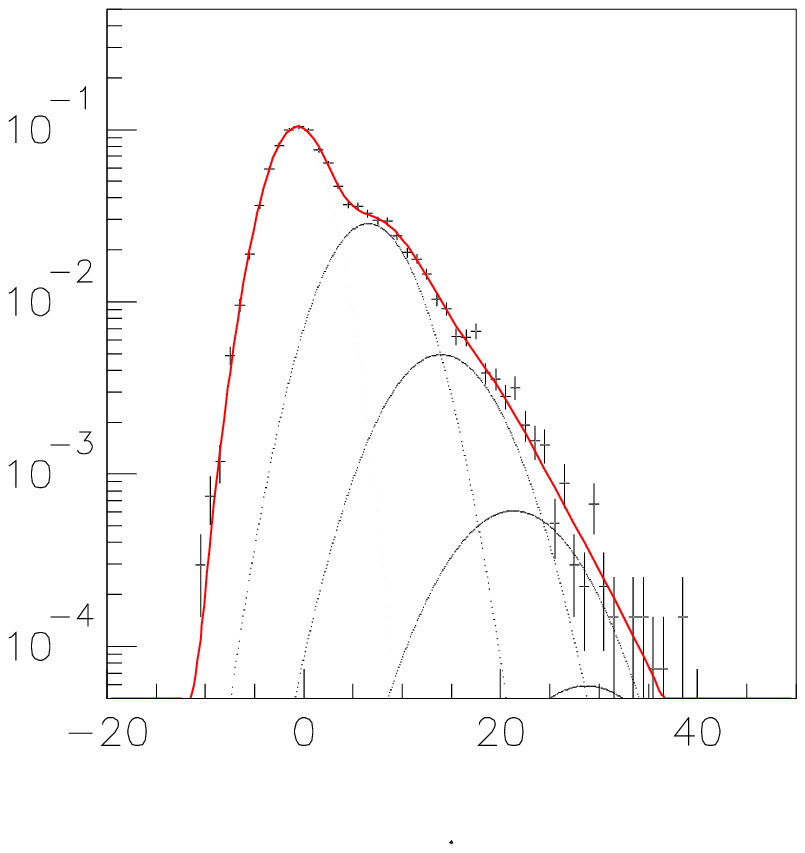}
}
\caption{
Left: Pulse-size (in arbitrary units) distribution from a single PMT 
made with events with LED light level 1 (one LED flashing).
The distribution has been fit with a five-parameter function. 
The parameters are: pedestal (zero photo-electron peak) mean and width,
single-photoelectron peak mean and width, and mean number of photoelectrons
per pulse. 
Right: As in the left-hand plot but made with LED light level 2 (two 
LEDs flashing simultaneously).
}
\label{spe_1_2}
\end{figure}

To see if this flasher can be used successfully for single-photoelectron 
studies we made a 120,000-event run (10 minutes at 200 Hz) with 
the neutral-density filter mounted on a VERITAS camera. 
We then divided the data into subsets according to how many LEDs were on in an 
event.
This is required since the underlying assumption of Poisson statistics relies 
on the distribution of light pulse intensities being essentially a delta 
function.
For a given light level we histogram the pulse sizes for each channel and fit
the distribution with the five-parameter function described earlier.
Sample plots with fits superimposed are shown in Figure~\ref{spe_1_2}.
The left plot is made with  data from events where one LED was illuminated and 
the right plot is made with data from events where two LEDs were illuminated.

\begin{figure}[]
\centerline{\includegraphics[width=1.2\textwidth]{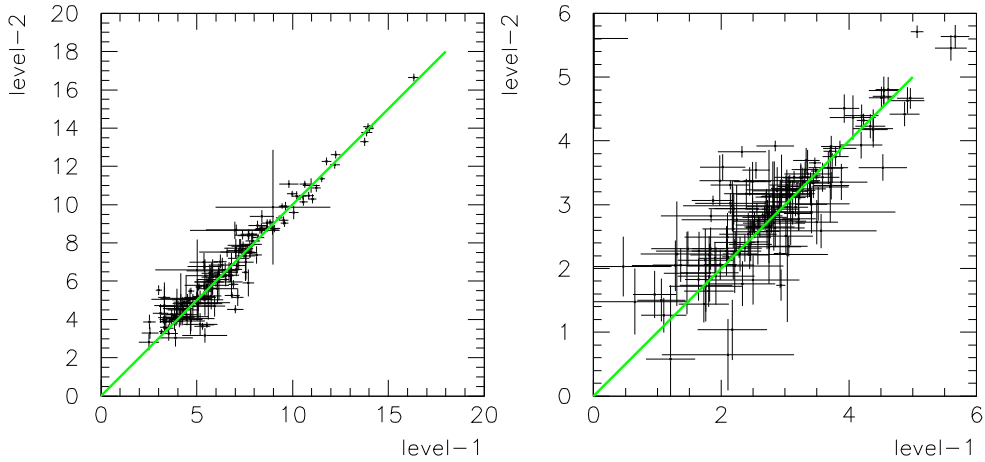}}
\vspace*{0.0cm}
\caption{Comparison of the means (left) and widths (right)
of the single-photoelectron peaks
made with one LED illuminated (level 1) and two LEDs illuminated (level 2).
The solid lines are to guide the eye and are not fits.
All scales are in arbitrary units.
}
\label{spe_summary}
\end{figure}

As a check on the robustness of the procedure we can compare the
single-photoelectron values obtained for the two different light levels.
Changing the light level should only change the relative numbers of 
zero, one, two, etc photoelectrons. 
The mean value of the charge distribution created by a single photoelectron 
should stay the same.
This is demonstrated in Figure~\ref{spe_summary} where the 
single-photoelectron means ($\mu_{spe}$) and standard deviations ($\sigma_{spe}$)
for the two light levels are
plotted against each other and a strong correlation is observed.

\begin{figure}[]
\centerline{\includegraphics[width=1.1\textwidth]{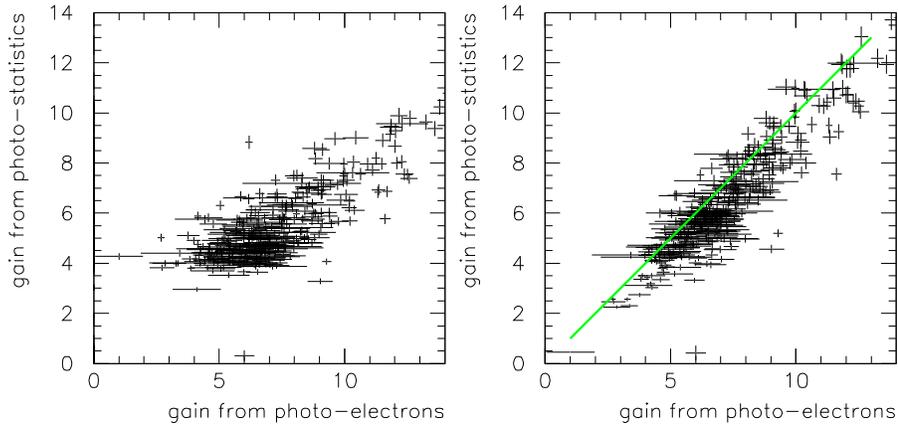}}
\vspace*{0.0cm}
\caption{Comparison of the gains determined using single photoelectrons (x-axis)
and photostatistics (y-axis).
In the left plot the raw results are used. 
In the right plot the $\sigma^2/\mu$ values have been multiplied by the Polya 
correction factor discussed the text ($1/(1+\alpha^2)$ where 
$\alpha=\sigma_{spe}/\mu_{spe}$) as well as the factor $1.05^{7.46}$ to account
for the boosted high voltage (5\% higher) used for the single-photoelectron run.
The solid line is to guide the eye and is not a fit.
}
\label{pe_vs_stats}
\end{figure}

\begin{figure}[]
\centerline{\includegraphics[width=0.9\textwidth]{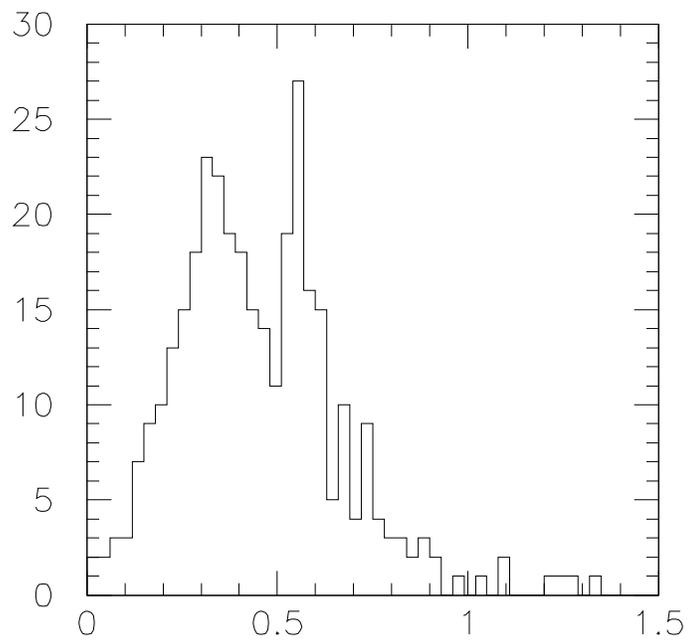}}
\vspace*{0.0cm}
\caption{Distribution of the Polya correction factor 
($1/(1+\alpha^2)$ where $\alpha=\sigma_{spe}/\mu_{spe}$)
discussed in the text.
}
\label{alpha}
\end{figure}

A final check on the procedures can be made by plotting the gain determined 
using photostatistics (the slope of the $\sigma^2$ vs $\mu$ plots, as in 
Figure~\ref{led_slopes}) against the gain from the single-photoelectron method.
This is done in Figure~\ref{pe_vs_stats}. 
The left plot shows uncorrected data and the right plot shows the data
after the photostatistics values have been been adjusted by two factors.
The first factor is the Polya correction referred to in a previous section. 
This is 
($1/(1+\alpha^2)$ where $\alpha=\sigma_{spe}/\mu_{spe}$) and $\sigma_{spe}$
is the width of the single-photoelectron peak while $\mu_{spe}$ is its mean.
The distribution of $\alpha$ is shown in Figure~\ref{alpha}.
The other factor accounts for the increase in gain due
the increased high-voltage (5\% above nominal) used for the single-photoelectron
run in order to increase the separation of the single-photoelectron peak 
from the pedestal.
The PMT gains have a power-law dependence on high-voltage: 
$G = G_0 (HV/HV_0)^{7.46}$.

\begin{figure}[]
\centerline{\includegraphics[width=0.9\textwidth]{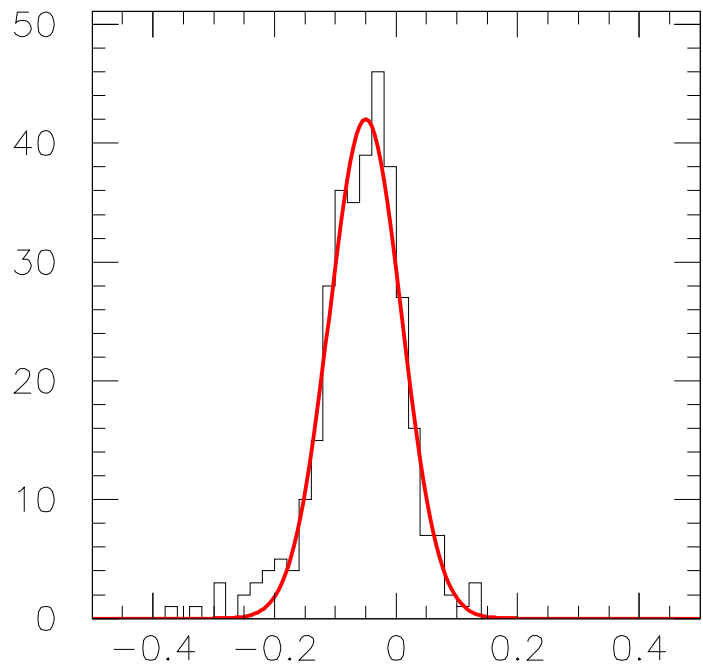}}
\vspace*{0.0cm}
\caption{
Ratio of the difference between the two gain estimators in 
Figure~\ref{pe_vs_stats} to their sum.
The fitted Gaussian has a mean of -0.05 and standard deviation of
0.06.
}
\label{stat_pe_ratio}
\end{figure}

As can be seen from the figure, the two estimators are correlated but there
is a systematic shift from exact correlation.
This can also be seen in Figure~\ref{stat_pe_ratio} where the ratios of the 
differences of the two estimators to their sums are plotted. The mean is 
at -5\% and the standard deviation is 6\%.
More work needs to be done to understand this discrepancy but is beyond the 
scope of this paper.

Note that entries in Figures~\ref{spe_summary}, \ref{pe_vs_stats}, 
and~\ref{alpha} rely on fits to pulse-size distributions.
All channels where the fits converged and had a $\chi^2_{dof}$ less than 3.0 
are included in the plots.

\section{Conclusions}

We have developed an
LED-based flasher system that can be used for all of the tasks
currently assigned to the VERITAS laser system.
In addition it can be run at much 
higher rates so calibration runs can be carried
out more quickly or can benefit from improved statistics for the same length of
run.

The LED flasher is also much less expensive than a laser and the LED system
is inherently a distributed one.
A separate light source is installed on each telescope of the array.
This could make it an attractive option for future projects such as 
AGIS~\cite{agis} and CTA~\cite{cta} where a large number of telescopes 
distributed over a large area would make 
a system based on a central laser sending light through optical fibres 
impractical.

\section{Acknowledgements}

This work has been supported by the Natural Sciences and Engineering Research 
Council (NSERC).
We are grateful to our colleagues in the VERITAS collaboration for the use of
components of the detector in evaluating our LED system.

\end{document}